\documentclass{IEEEtran}  

\usepackage[utf8]{inputenc}
\usepackage{amssymb}
\usepackage{amsmath}
\usepackage{color}
\usepackage{longtable}
\usepackage{graphicx}

\newcommand{\Adv}{\ensuremath{\mathcal{A}}}
\newcommand{\Sim}{\ensuremath{\mathcal{S}}}
\newcommand{\Zq}{\mathbb{Z}_Q}
\newcommand{\FFSM}{\ensuremath{\mathcal{F}_{\mathsf{FSM}}}}

\newtheorem{thm}{Theorem}[section]
\newenvironment{theorem}{\begin{thm}}{\end{thm}}

\def\suchthatt{\: :\:}
\def\getsr{\stackrel{{\scriptscriptstyle\$}}{\leftarrow}}

\begin{document}
 
\title{Round and Communication Balanced Protocols for Oblivious Evaluation of Finite State Machines}

\author{Rafael~Dowsley, Caleb~Horst, Anderson~C.~A.~Nascimento

\thanks{Caleb Horst and Anderson~C.~A.~Nascimento are with the Institute of Technology, University of Washington Tacoma. E-mails: \{calebjh, andclay\}@uw.edu.}
\thanks{Rafael Dowsley is with the Faculty of Information Technology, Monash University, Australia. Email: rafael.dowsley@monash.edu.}
}

\maketitle

\begin{abstract}
We propose protocols for obliviously evaluating finite-state machines, i.e., the evaluation is shared between the provider of the finite-state machine and the provider of the input string in such a manner that neither party
learns the other's input, and the states being visited are hidden from both. For alphabet size $|\Sigma|$, number of states $|Q|$, and input length $n$, previous solutions have either required a number of rounds linear in $n$ or communication $\Omega(n|\Sigma||Q|\log|Q|)$. Our solutions require 2 rounds with communication $O(n(|\Sigma|+|Q|\log|Q|))$. We present two different solutions to this problem, a two-party one and a setting with an untrusted but non-colluding helper. 
\end{abstract}


\IEEEpeerreviewmaketitle

\section{Introduction}

Finite-state machines (FSM) are a simple, but useful computational model. The ability to evaluate a finite-state machine in a privacy-preserving way allows for some interesting use cases. Finite-state transducers (which are easily extended from automata) can be used to count occurrences of multiple strings or string types in an input text. This allows for uses such as pre-processing text for machine learning algorithms by extracting linguistic features (e.g. LIWC\footnote{https://liwc.wpengine.com/}), or comparing an input text against some text of interest. Many of the previous works on oblivious FSM evaluation have been targeted at the application of DNA sequence matching. 

This work presents a pair of protocols that allow two parties, one holding a private FSM and the other holding a private input string, to determine the outputs of that FSM on the input string while revealing minimal information (namely the sizes) about the private inputs. Compared to previous solutions, our protocols achieve a better balance of round and communication complexities as the input alphabet grows in size. One existing line of solutions \cite{frikken2009practical} allows to obtain low constant round complexity at the cost of having 
$\Omega(n|\Sigma||Q|\log|Q|)$ communication complexity (where $\Sigma$ is the input alphabet, $n$ is the input length, and $Q$ is the set of states). Another line of solutions \cite{Blanton2010} allowed very low communication, potentially as low as $O(n\log^2(|Q||\Sigma|))$, at the cost of $n$ rounds and very heavy computation.
Our proposition is a 2-round protocol that has communication $O(n(|\Sigma|+|Q|\log|Q|))$. We present a two-party protocol and also a way to make the protocol even more efficient if an additional untrusted but non-colluding server is available to help in the computation. As the input alphabet grows, the communication complexity of our protocol is much better than the previous constant round protocols, therefore for applications where the alphabet is significant (such as text input processing) the communication is significantly improved.

\subsection{Related Works}

Several protocols have been previously developed to address the oblivious evaluation of FSM, starting with Tronosco-Pastoriza et al. \cite{troncoso2007privacy} that targeted the application of DNA sequence matching and used basic secret sharing with homomorphic encryption (HE) and oblivious transfer (OT) to evaluate the FSM one input symbol at a time. They considered the evaluation being split between two parties, a server and a client. The client holds $x \in \Sigma^n$ and the server holds an FSM. At each point in the evaluation, the parties hold an additive secret sharing representing $q_i\in Q$, the $i$-th state. To retrieve shares of the next state, the server first masks all the entries of the transition matrix with a round-specific random value $r$, then rotates the entire matrix by his share of the previous state. The client uses an additively HE scheme (e.g. Paillier) to encrypt a unary encoding of his share, i.e., $v_1,...v_{|Q|}$ with $v_i$ being encryptions of the value 0 except for the index of his state share that is an encryption of the value 1. Using the homomorphic properties, the server multiplies his matrix by this vector and attains an encrypted vector representing the transition function for the current state. The server and client then execute a $1$-out-of-$|\Sigma|$ OT protocol to retrieve the $x_i$'th element (the encrypted entry corresponding to the client's input), which the client can decrypt to learn his share of the next state, while the server keeps the value $r$ as his share. This process is repeated until the entire input $x$ is processed.\footnote{As pointed out by Tronosco-Pastoriza et al., the order can be switched by working on the transpose of the matrix, encrypting the input as a vector, and then doing a $1$-out-of-$|Q|$ OT, depending on which is more efficient.}
Blanton and Aliasgard \cite{Blanton2010} slightly improved on this protocol to use only OT and improve the communication complexity, as well as demonstrating how evaluation can be outsourced to offload the computational costs from the input and FSM holders. Instead of using the unary vector of HE encryptions to select the correct vector then doing OT on that vector, they perform OT on the entire matrix by representing it as a vector then fetching the proper index (as in 2D array indexing). This incurs a $1$-out-of-$|\Sigma||Q|$ OT per round instead of $|Q|$ Paillier and a $1$-out-of$|\Sigma|$ OT. By using a bandwidth optimized protocol (such as the one in \cite{unal2014bandwidth}) in lieu of the trivial HE, this approach can achieve the lowest communication complexity of any known protocol, at the cost of being the most computationally expensive. The downside of both of these protocols is that they all suffer from a heavy computational load due to the large amount of HE or OT called for, commonly requiring expensive public-key encryption operations, and are restricted to running in no fewer than $n$ rounds.

Frikken \cite{frikken2009practical} introduced a different approach that generalizes the idea behind garbled circuits in order to obtain an oblivious FSM protocol requiring only two rounds of communication. Additionally, this technique requires only $n$ OTs, and is otherwise composed entirely of fast symmetric key operations. However, it has a high communication overhead, as the entire FSM is sent in garbled form $n$ times. Mohassel et al. \cite{Mohassel2012} reformatted the FSM as using binary inputs and obtained better efficiency. The idea of this approach is that the entire transition matrix is garbled $n$ times, once for each input character. For each matrix, first a permutation (rotation) is applied to the set of next states to hide the order. Then, for each state and input, the next state is encoded in such a manner that it can be uncovered only using both the key for the current state (gained as the FSM is evaluated) and the key for the specific input (retrieved using OT at the start of the evaluation). To perform the evaluation, the data holder uses $1$-out-of-$|\Sigma|$ OT to retrieve the keys corresponding to each symbol in his input and the FSM holder sends the garbled FSM to the data holder. From the initial state (which can be made public due to the garbling) and the input keys, the data holder evaluates the FSM by ungarbling the correct entry for the current state and input to determine the next state and next state key until the final state is reached. 

A few more approaches have been used for oblivious FSM evaluation: Laud and Willemsen \cite{laud2013universally} utilize an  arithmetic black box, an abstraction of distributed computation  such as secret sharing, and a polynomial representation of the transition function to achieve quite low `online' communication complexity, however the round complexity is still tied to the  length of the input, and the preprocessing still requires a large amount of communication. Other works further use additional permutations and matrix multiplication to complete Deterministic Finite Automaton (DFA) and Nondeterministic Finite Automaton (NFA) evaluations \cite{Wei2012,Wei2015,sasakawa2014oblivious}, sometimes even working over encrypted inputs. 
In general these protocols are less efficient.

Oblivious FSM can be very useful in the pre-processing phase of privacy-preserving machine learning solutions. Recent advances in privacy-preserving machine learning based on secure multi-party computation focused mostly on the training and scoring phases of models such as: linear regression \cite{AISec:CDNN15,mohassel2017secureml,GasconSB0DZE17,IEEENSRE:ADMW+19}, logistic regression \cite{mohassel2017secureml,IEEETDSC:CDHK+17,idash}, neural networks \cite{mohassel2017secureml,liu2017oblivious,wagh2019securenn,dalskov2019secure,agrawal2019quotient,riazi2019xonn,bittner2021private,kumar2020cryptflow,cryptflow2,pentyala2021privacypreserving} and decision trees (ensembles) \cite{HooghSCA14,IEEETDSC:CDHK+17,fritchman2018privacy,escudero2020,PETS:ACCDMN+22}. Only few works considered the privacy-preserving execution of the (extremely important) data pre-processing stage, obtaining solutions for privacy-preserving feature extraction \cite{NeurIPS2019,resende2021fast} and privacy-preserving feature selection \cite{li2021privacypreserving}.

\subsection{Our Techniques}

We improve the performance of previous results in one key way. Instead of completing each state transition in sequence using OT, or garbling and sending the entire transition matrices, we garble the transition matrices and then select only the relevant columns to be transmitted.  This allows for a constant number of rounds, as well as communication $O(n(|\Sigma|+|Q|\log|Q|))$.

Our protocol is based on garbling schemes. We select $n|Q|$ uniformly random keys $k_{q}^i$ of length $\kappa$ to represent the keys for each of the $Q$ states in each of the $n$ garbled transition matrices, as well as $n$ uniformly random values $r_i\in \Zq$ which are used to rotate the transition matrices. For a transition function $\delta: (Q, \Sigma) \mapsto Q$, which can be represented as a $|Q|\times|\Sigma|$ matrix $\Delta$, and for all $i\in\{1,\ldots,n\}, q\in\{1,\ldots,|Q|\}, \sigma\in\{1,\ldots,|\Sigma|\}$, the garbled form of each entry then becomes
\[
G_{\mathsf{fsm}}(q,\sigma, i) := 
H(k_{q}^i || \sigma) \oplus (\Delta_{q^\prime,\sigma}+r_{i+1} ||  k^{i+1}_{\Delta_{q^\prime,\sigma}+r_{i+1}}),
\]
where $q^\prime = q-r_i$ is the true state
and $H: \{0,1\}^{\kappa+\log|\Sigma|}\mapsto\{0,1\}^{\kappa+\log|Q|}$ is a hash function. This garbled form is an encryption of the next permuted state for the given permuted state and input, along with the corresponding key so that it can be decrypted to continue the evaluation. Any IND-CPA secure encryption scheme could be used, but for efficiency reasons we choose to use this encryption with a hash function and assume the random oracle model. Note that this garbling does not use keys related to the input string (as in previous solutions), but the careful usage of OT enables the design of a secure protocol. Given the current state's key and the ciphertext, it is clearly possible to decrypt the permuted state to transition to and the corresponding key. The key point is that the careful usage of OT allows the design of a transferring mechanism in which the client only obtains the ciphertexts corresponding to his input string. As for the output, there are many possibilities depending on the applications: sometimes it should be revealed to the client, other times to the server, and yet in other scenarios it should be kept secret shared between the client and the server so that none of them individually know the output and it can be used in further secure computations. In the description of our protocols we will keep the output secret shared since this is more general and it can be easily modified in order to open the output to one of the parties by just having the other party forwarding its share. In this type of output, the server picks a random value $r$ (which will be his share of the output) and then the output is included in the garbled FSM by replacing the final transitions with $1 + r$ if the final state that would be transitioned to is accepted and 
$r$ otherwise.

Once the matrices are garbled, the client performs a selection operation to retrieve the columns corresponding to his inputs. \footnote{Technically, for the first transition, only the transitions from $q_0$ need to be garbled, not the entirety of $\Delta$, and the only entry of interest is the transition from $q_0$ on the first symbol of $x$.} This is done in parallel for all garbled FSM. He then proceeds to the evaluation of the FSM. First, the server must provide the state key for the first transition, i.e., the key $k_{q_{\mathsf{Init}}}^1$ corresponding to the initial state of the garbled FSM. Given $k_{q_{\mathsf{Init}}}^1$, the client can clearly use it to decrypt the next state and the corresponding key. Using that information, he can iterate through the columns until the evaluation is complete.

\section{Preliminaries}\label{sec:pre}

\subsection{Finite-State Machines}

A finite-state machine (FSM) is a simple model of computation and a basic FSM is defined to be a tuple 
\[\mathcal{M}=(Q, q_0, \Sigma, \delta, Q_{\mathsf{accept}}),\] 
where $Q$ is a set of states, $q_0\in Q$ is the initial state, $\Sigma$ is the alphabet in use, $\delta$ is a transition function $\delta: (Q, \Sigma) \mapsto Q$ and $Q_{\mathsf{accept}} \subset Q$ is the set of accept states. To mathematically represent $\delta$, a matrix $\Delta$ is often used. $\Delta$ is a $|Q|\times|\Sigma|$ matrix where each entry has $\log_2{|Q|}$ bits and represents the state transitioned to for each state/input pair. An input $x\in\Sigma^n$ is \textit{accepted} on $\mathcal{M}$ if the repeated application of $\delta$ from $q_0$ using the input $x$ results in a final transition to a state in $Q_{\mathsf{accept}}$. We can consider $\mathcal{M}$ as having an output of $1$ for accepted strings, and an output of $0$ for any other string.

A more versatile version of the FSM is the finite-state transducer (FST), in which $Q_{\mathsf{accept}}$ is replaced by $\Gamma$, an output alphabet (usually $\mathbb{Z}/m\mathbb{Z}$ for some $m$) and $\lambda$, an output rule. The two common FST models are Moore machines and Mealy machines. In a Moore machine, we have $\lambda: Q \mapsto \Gamma^\rho$, and in a Mealy machine $\lambda: Q\times\Sigma \mapsto\Gamma^\rho$, where $\rho$ represents the number of output variables. We can represent this readily as a matrix $\Lambda$ of either dimension $1\times |Q|$ or $|\Sigma|\times|Q|$, respectively, where each entry is $\rho\log_2{|\Gamma|}$ bits in size. This can also be represented by appending the outputs to the entries of $\Delta$, however doing so will result in redundancy in a Moore machine.

\subsection{Homomorphic Encryption}

Homomorphic encryption (HE) allows one to compute functions on encrypted data that will be reflected in the decryption. In this work we will consider 
additively homomorphic encryption schemes that are IND-CPA secure \cite{GoldwasserM84}. In such schemes, for a public/secret-key pair $(pk,sk)$ and for ciphertexts $\mathsf{Enc}_{pk}(x)$ and $\mathsf{Enc}_{pk}(y)$, there exists an efficiently computable operation denoted by $*$ such that $\mathsf{Dec}_{sk}(\mathsf{Enc}_{pk}(x)*\mathsf{Enc}_{pk}(y))=x+y$. In other words, the encryption scheme allows the efficient addition of enciphered values. Furthermore, we will require that given a constant $c$ and $\mathsf{Enc}_{pk}(x)$, there is an efficient way of computing a ciphertext corresponding to $cx$ and to $c+x$. One example of a encryption scheme meeting all those requirements is Paillier's encryption scheme \cite{paillier1999public}. 

\subsection{Additive Secret Sharing}

For an integer $q$ and a value $x \in \mathbb{Z}_q$, an additive secret sharing of $x$ between two parties is created by picking uniformly random $x_1, x_2 \in \mathbb{Z}_q$ constraint to $x_1+x_2 = x \mod q$ and delivering $x_1$ to the first party and $x_2$ to the second party. Note that any single party $P_i$ does not learn any information about $x$ from its share $x_i$, but $x_1$ and $x_2$ can be recombined to obtain $x$.

\subsection{Security Model}

We use the same half-simulation security model as Mohassel et al. \cite{Mohassel2012}, which is briefly explained here. Please refer to \cite{goldreich2009foundations} for further details about security models.

Let $f = (f_1,f_2)$ of the form $f: \{0,1\}^* \times \{0,1\}^* \rightarrow \{0,1\}^* \times \{0,1\}^*$ be a two party computation and $\pi$ be a two-party protocol for computing $f$ between the parties $p_1$ with input $x$ and $p_2$ with input $y$.

Full-security (simulation-based security) is defined by requiring indistinguishability between a real execution of the protocol $\pi$ and an ideal execution in which there is a trusted third party (TTP) who receives the parties input, evaluates the function and outputs the results to them. An admissible adversary is one that corrupts exactly one of the two party. $\mathcal{A}$ also knows an auxiliary input z. Without loss of generality we assume that $\mathcal{A}$ corrupts the first party. 
If the adversary is semi-honest, he follows the protocol instructions but can try to obtain additional information. On the other hand, if the adversary is malicious he can deviate from the protocol instructions. 

In the real world, the honest party follows the  instructions of protocol $\pi$ and responds to messages sent by $\mathcal{A}$ on behalf of the other party. Let $\mathsf{View}_{\pi, \mathcal{A}}(x,y)$ denote $\mathcal{A}$’s view through this interaction, and let $\mathsf{Out}_{\pi, \mathcal{A}}(x,y)$ denote the output of the honest party. The execution of $\pi$ in the real model on input pair $(x, y)$ is defined as follows:
$$\mathsf{Real}_{\pi, \mathcal{A}(z)}(x,y)=(\mathsf{View}_{\pi, \mathcal{A}}(x,y), \mathsf{Out}_{\pi, \mathcal{A}}(x,y)).$$

In the ideal model, in which there is a TTP, the honest party always sends its input $y$ to TTP, while the malicious party can send an arbitrary input $x'$. The TTP first replies to the first party with $f_1(x',y)$ (in case it receives only one valid input, the trusted party replies to both parties with a special symbol $\bot$). In case the first party is malicious it may, depending on its input and the trusted party’s answer, decide to stop the trusted party by sending it $\bot$ after receiving its output. In this case the trusted party sends $\bot$ to the honest party. Otherwise (i.e., if not stopped), the trusted party sends $f_2(x', y)$ to the honest party. The honest party outputs whatever is sent by the trusted party, and $\mathcal{A}$ outputs an arbitrary function of its view. Let $\mathsf{Out}_{f, \mathcal{A}}(x,y)$ and $\mathsf{Out}_{f}(x,y)$
denote the output of $\mathcal{A}$ and the honest party respectively in the ideal model. The execution of $\pi$ in the ideal model on input pair $(x, y)$ is defined as follow:
$$\mathsf{Ideal}_{f, \mathcal{A}(z)}(x,y)=(\mathsf{Out}_{f, \mathcal{A}}(x,y), \mathsf{Out}_{f}(x,y)).$$

Protocol $\pi$ securely computes $f$ in the presence of static adversaries if for every  admissible non-uniform probabilistic polynomial-time adversary $\mathcal{A}$ in the real model, there exists an admissible  
nonuniform probabilistic expected polynomial-time adversary $\mathcal{S}$ in the ideal model, such that the distributions of the real and ideal executions are indistinguishable:
$$\{\mathsf{Real}_{\pi, \mathcal{A}(z)}(x,y)\}\approx\{\mathsf{Ideal}_{f, \mathcal{S}(z)}(x,y)\}.$$

We will refer to the two parties based on their input to the protocol. The \textit{Client} provides an input string to be run through the FSM provided by the \textit{Server}.

The functionality for computing the finite-state machine $\FFSM$ waits for the server to input a FSM
$\mathcal{M}=(Q, q_0, \Sigma, \delta, Q_{\mathsf{accept}})$ and the client to input a string $x=x_1\ldots x_n\in \Sigma^n$. It is assumed that $|Q|$, $\Sigma$ and $n$ are public. $\FFSM$ computes the result of evaluating $\mathcal{M}$ on input string $x$ and secret shares the result between the \textit{Client} and the \textit{Server}.

In our protocols, we achieve a weaker notion of security against a malicious \textit{Server}. This notion intuitively guarantees that a corrupted party will not learn any information about the input of the honest party (however, this does not  guarantee that the parties joint outputs in the real world is simulatable in an ideal world).  Without loss of generality we assume that the first party is the malicious one. We say that protocol $\pi$ is private against a malicious party $p_1$ if the advantage of any non-uniform polynomial-time adversary $\mathcal{A}$ corrupting $p_1$ in the real world is negligible in the following game:
\begin{enumerate}
    \item $\mathcal{A}$ is given $1^\lambda$ and generates $y_0, y_1 \in \{0, 1\}^n$ for some positive integer $n$. $\mathcal{A}$ sends $y_0, y_1$ to $p_2$.
    \item $p_2$ generates a random bit $b \getsr \{0, 1\}$ and uses $y_b$ as his input in protocol $\pi$.
    \item At the end of the protocol $\pi$, $\mathcal{A}$ outputs a bit $b'$. $\mathcal{A}$'s advantage in the game is defined as $\Pr(b=b')-1/2$.
\end{enumerate}

In our protocol with three parties, to achieve a higher efficiency, we allow for a third \textit{Helper} party that is expected to not collude with either of the other two parties, but is not
trusted to see any input.

\subsection{Oblivious Transfer Protocol}

In this work we use as a building block an oblivious transfer (OT) protocol in which the sender Alice has inputs $b_1, \ldots, b_t$ such that $|b_1|=\ldots= |b_t|=\ell$ and the receiver Bob chooses the index $u \in \{1,\ldots, t\}$ of the input $b_u$ that he wants to receive. In order to achieve the desired communication complexity of $O(n(|\Sigma|+|Q|\log|Q|))$ for our oblivious FSM protocol, the OT protocol needs to have a communication complexity of $O(\ell+t)$. We will use the following OT protocol that is secure against a semi-honest receiver Bob:

\begin{enumerate}
    \item Bob represents its choice $u$ using an one-hot encoding, i.e., a binary vector $(v_1, \ldots, v_t)$  such that only $v_u=1$. Bob then encrypts each $v_i$ using the additively homomorphic encryption scheme with his own public-key $pk$ to obtain $\mathsf{Enc}_{pk}(v_i)$. He sends all ciphertexts to Alice.
    \item Alice multiply each $\mathsf{Enc}_{pk}(v_i)$ by $b_i$ and sum all the results to obtain the ciphertext $d$ which corresponds to the plaintext $\sum_{i=1}^t v_i b_i=b_u$. In order to re-randomize the ciphertext, Alice encrypts $0$ using $pk$ and uniform randomness and adds it to $d$ to obtain $d'$ that also encrypts $b_u$. Alice sends the ciphertext $d'$ to Bob. 
    \item Bob uses his secret key to decrypt $d'$ and obtain the chosen input $b_u$.
\end{enumerate}

\section{Protocol}

We consider the setting in which the server has as input a FSM $\mathcal{M}=(Q, q_0, \Sigma, \delta, Q_{\mathsf{accept}})$, the client has as input a string $x=x_1\ldots x_n\in \Sigma^n$. It is assumed that the client knows $|Q|$ and $\Sigma$, while the server knows $n$.

Similarly to \cite{frikken2009practical,Mohassel2012}, we will garble $n$ times the state transition matrix $\Delta$. Each garbling consists of a permutation and symmetric encryptions.
The suggested permutation is adding a random mask value modulo $|Q|$. We will use $r_i$ for these rotation values, and forgo including the modular reduction notation.
The server chooses $n\cdot|Q|$ random keys denoted $k^i_q$ of length $\kappa$, one for each state in each of the $n$ matrices. 
Let $q^\prime = q-r_i$ denote the true state. For all $i\in\{1,\ldots,n\}, q\in\{1,\ldots,|Q|\}, \sigma\in\{1,\ldots,|\Sigma|\}$, the garbled form of each entry is
\[
G_{\mathsf{fsm}}(q,\sigma, i) := 
H(k_{q}^i || \sigma) \oplus (\Delta_{q^\prime,\sigma}+r_{i+1} ||  k^{i+1}_{\Delta_{q^\prime,\sigma}+r_{i+1}}),
\]
where $H: \{0,1\}^{\kappa+\log|\Sigma|}\mapsto\{0,1\}^{\kappa+\log|Q|}$ is a hash function. Later on this section we will address more thoroughly how the output can be included in the garbled FSM, for now, we simply let the server hold a random masking value $r$ and 
replace the final transition values in the garbled FSM with the value $1 + r$ if the final state is accepted and $r$ otherwise.

By combining this garbled FSM scheme with mechanisms that obliviously and efficiently transmit only the necessary columns, we can make the $|\Sigma|$ and $|Q|\log|Q|$ communication factors additive instead of multiplicative while having only a constant number of rounds. 

Once the matrices are garbled, the server and the client execute $n$ instances of the OT protocol specified in Section \ref{sec:pre} in order to transmit the necessary columns of the garbled transition matrices. In each instance, the server's inputs are the columns of one garbled transition matrix and the client uses his input string $x$ to select one column from that garbled transition matrix that he will receive. This selection is similar to what happens in the original sequential protocol \cite{troncoso2007privacy}, 
however all column selections occur at once and all selected columns are sent at once, 
as opposed to completing each state selection before proceeding to the next round. The initial key $k_{q_{\mathsf{Init}}}^1$ is also sent to the client, who is then able to decrypt the relevant garbled entries and use these to evaluate the FSM and retrieve the output. The protocol $\pi$ proceeds as follows:

\begin{enumerate}
\item The server prepares the garbled transition matrices $G_{\mathsf{fsm}}(q,\sigma, i)$.

\item The server and the client execute in parallel $n$ instances of the $1$-out-of-$|\Sigma|$ OT protocol specified in Section \ref{sec:pre} in which in the $i$-th instance the server inputs are the columns of the $i$-th garbled transition matrix $G_{\mathsf{fsm}}(\cdot,\cdot, i)$ and the client selection is $x_i$.

\item The server sends the key $k_{q_{\mathsf{Init}}}^1$ corresponding to the initial state of the garbled FSM to the client (note that this can also be done in parallel with the previous step).

\item The client decrypts the first garbled transition vector and ungarbles the first state transition using the given key. He then iteratively evaluates the FSM by decrypting and ungarbling the entry of the next vector 
using the ungarbled key.
\end{enumerate}

\subsection{Security Proof}

\begin{theorem}
The protocol $\pi$ is secure against a semi-honest client and private against a malicious server.
\end{theorem}

\textbf{Security against a Semi-Honest Client:}
The adversary $\Adv$ controls the client in the real world. In the ideal world the simulator $\Sim$ runs internally a copy of the adversary $\Adv$ and simulates an execution of the protocol $\pi$ for $\Adv$. 

$\Sim$ gives the client's input $x=x_1\ldots x_n$ to $\FFSM$ and gets the client share $z \in \{r,r+1\}$ of the $\FFSM$'s output (for a random masking value $r$). $\Sim$ proceeds in the following way to create the relevant columns of the garbled FSM in the internal simulation of the execution of protocol $\pi$: 

\begin{enumerate}
    \item $\Sim$ chooses $n\cdot|Q|$ random keys of length $\kappa$ denoted by $k^i_q$, for $i\in\{1,\ldots,n\}, q\in\{1,\ldots,|Q|\}$.
    \item For $i=1,\ldots,n$, $\Sim$ samples an uniformly random value $q_i$ from $\{1,\ldots,|Q|\}$.
    \item For $i=1,\ldots,n-1$, $\Sim$ fixes 
\[
G_{\mathsf{fsm}}(q_i,x_i, i) :=
H(k_{q_i}^i || x_i) \oplus \left(q_{i+1}||  k^{i+1}_{q_{i+1}}\right),
\]
and it also fixes \[
G_{\mathsf{fsm}}(q_n,x_n, n) := H(k_{q_n}^n || x_n) \oplus z,
\]
so that the path that will be traversed by the client results in the desired share of the output.
    \item For $i=1,\ldots,n$, and all $q \in \{1,\ldots,|Q|\} \setminus \{q_i\}$, $\Sim$ fix 
\[
G_{\mathsf{fsm}}(q,x_i, i) := H(k_{q}^i || x_i) \oplus
\left( 0^{\kappa+\log|Q|} \right).
\]
\end{enumerate}
The simulator $\Sim$ then encrypts the columns
$G_{\mathsf{fsm}}(\cdot,x_i, i)$ for $i=1,\ldots,n$ using the public-key $pk$ of the client, and sends to the simulated adversary $\Adv$ the ciphertexts and the key $k_{q_1}^1$ corresponding to the initial state as being the message of the honest server in the simulated execution of Protocol $\pi$. The simulator outputs the final view of the simulated adversary $\Adv$. 

The security against a semi-honest client follows trivially from the IND-CPA security of the symmetric encryption scheme that is used in the garbled transition matrices. A classical hybrid argument can be employed to replace one by one the encryptions of zero used by the simulator by the encryptions of the real garbled values used in the real world, thus proving that the real and ideal worlds are indistinguishable.

Protocol $\pi$ can be made secure against malicious clients by adding zero-knowledge proofs to prove that in each instance of the OT protocol only one entry of the one-hot encoding sent by the client is equal to 1.

\textbf{Privacy against a Malicious Server:} The only messages sent by the client in Protocol $\pi$ are for performing the OTs, therefore from the IND-CPA security of the additively homomorphic encryption scheme used in the OTs, it follows trivially that Protocol $\pi$ is private against a malicious server.

\subsection{Output and Transducing}
To generate output, we can consider two ways to incorporate the output function
into the garbling scheme. The first being the one stated before, which applies 
only to the basic finite state automaton.
On the last state transition, we replace the next state value with the 
acceptance value $0/1$ (masked with a random $r$ that is known by the server):
\[
G_{\mathsf{fsm}}(q,\sigma, n) := 
\left\{
\begin{array}{ll}
H(k_{q}^n || \sigma) \oplus r, \text{ if } \delta(q^\prime, \sigma) \notin Q_{accept}\\
H(k_{q}^n || \sigma) \oplus (r+1), \text{otherwise}
\end{array} 
\right.
\]

Note that this can be readily extended to non-binary outputs of a function $\lambda$,
\[
G_{\mathsf{fsm}}(q,\sigma, n) := 
H(k_{q}^n || \sigma) \oplus (r + \lambda(\delta(q^\prime, \sigma)).
\]

Furthermore, we can extend the protocol to transducers by adding this into the transitions rules, even if the output rule $\lambda$ outputs a number of separate outputs,
i.e. $\lambda: (Q, \Sigma) \mapsto \Gamma^\rho$.
We can use separate random masks $z_i$ for each $i\in \{1,\ldots,n\}$ and add the additional information in the garbled transition matrices, so that $G_{\mathsf{fsm}}(q,\sigma, i)$ now becomes 
\[
H(k_{q}^i || \sigma) \oplus (\Delta_{q^\prime,\sigma}+r_{i+1} ||  k^{i+1}_{\Delta_{q^\prime,\sigma}+r_{i+1}}||z_i + \lambda(\delta(q^\prime, \sigma))).
\]
 
In Moore's FST, where the outputs depend
only on the states and not on the inputs, we can cut the communication slightly by encrypting the outputs for each state using the state keys.

\section{Optimized Three-Party Protocol}

In the case that an untrusted, but non-colluding third party helper is available we can optimize the protocol by eliminating the public-key operations.
We will denote this new protocol by $\pi_{\mathsf{Helper}}$. The main idea is that the client secret shares its input between the server and the helper. The server and the helper can then use the secret sharings to compute partial answers to client, who can then XOR the partial answers to obtain the necessary columns of the garbled transition matrices. As explained in the next paragraphs, the protocol guarantees that the helper can also not learn any information about the finite state machine and that the partial answers do not leak additional information to the client.

The server shares the garbled matrices 
\[
G_{\mathsf{fsm}}(q,\sigma, i) := 
H(k_{q}^i || \sigma) \oplus (\Delta_{q^\prime,\sigma}+r_{i+1} ||  k^{i+1}_{\Delta_{q^\prime,\sigma}+r_{i+1}}),
\]
with the helper. This will not compromise security as the helper does not know any of the keys $k_{q}^i$. This sharing can happen during the protocol execution or at any time in an offline phase involving only the server and helper. The server also shares a uniformly random key $k_{SH}$ with the helper.

Each symbol $x_i$ in the client's input $x$ is encoded using one-hot-encodings, i.e., $x_i$ is encoded as $(x_{i,1},\ldots,x_{i,\Sigma})$ where $x_{i,j}=1$ if $x_i=j$, and otherwise $x_{i,j}=0$. The client then generates additive secret sharings of the bits $x_{i,j}$ by picking uniformly random bits $x_{i,j}^S,x_{i,j}^H$ such that $x_{i,j}^S \oplus x_{i,j}^H=x_{i,j}$, and send them, respectively, to the server and the helper.

For each $i \in \{1, \ldots, n\}$, and for the garbled transition matrix $G_{\mathsf{fsm}}(q,\sigma, i)$,
the server computes the XOR of the columns with index $j$ such that $x_{i,j}^S=1$, obtaining as the result the column vector
\[
G_{\mathsf{fsm}}^S(\cdot,x_i, i) := 
\bigoplus_{j \in\Sigma \suchthatt x_{i,j}^S=1} 
G_{\mathsf{fsm}}(\cdot,j, i).
\]
Similarly, the helper obtains the column vector
\[
G_{\mathsf{fsm}}^H(\cdot,x_i, i) := 
\bigoplus_{j \in\Sigma \suchthatt x_{i,j}^H=1} 
G_{\mathsf{fsm}}(\cdot,j, i).
\]
Note that 
\[
G_{\mathsf{fsm}}(\cdot,x_i, i) =
G_{\mathsf{fsm}}^S(\cdot,x_i, i) \oplus G_{\mathsf{fsm}}^H(\cdot,x_i, i)
\]
is exactly the column of the garbled transition matrix that the client should get. If $x_{i,j}=0$, then either $x_{i,j}^S=x_{i,j}^H=0$ in which case the $j$-th column is XORed in neither $G_{\mathsf{fsm}}^S(\cdot,x_i, i) $ nor 
$G_{\mathsf{fsm}}^H(\cdot,x_i, i)$, or 
$x_{i,j}^S=x_{i,j}^H=1$ in which case the $j$-th column is XORed in both $G_{\mathsf{fsm}}^S(\cdot,x_i, i)$ and 
$G_{\mathsf{fsm}}^H(\cdot,x_i, i)$ (and so cancelled out in the final XOR). On other hand, for the single value of $j$ such that $x_{i,j}=1$, the column will only be XORed in one of the partial answers. In order to guarantee that the partial answers do not leak additional information to the client, before sending the partial answers the server and the helper mask them with a random value, which cancels out when both partial answers are XORed. For a hash function $H'$ whose output size matches the size of one column of the garbled transition matrix, the server sends
\[
G_{\mathsf{fsm}}^S(\cdot,x_i, i) \oplus H'(k_{SH} || i)
\]
to the client, while the helper sends 
\[
G_{\mathsf{fsm}}^H(\cdot,x_i, i) \oplus H'(k_{SH} || i)
\]
to the client. By XORing both partial answers, the client can recover the appropriate column of the garbled transition matrix. To enable the client to evaluate the FSM, the server supplies the client with the key $k_{q_{\mathsf{Init}}}^1$ corresponding to the initial state of the garbled FSM.

\subsection{Security Proof}

\textbf{Security against a Semi-Honest Client:}

The adversary $\Adv$ controls the client in the real world. In the ideal world the simulator $\Sim$ runs internally a copy of the adversary $\Adv$ and simulates an execution of the protocol $\pi_{\mathsf{Helper}}$ for $\Adv$. 

$\Sim$ gives the client's input $x=x_1\ldots x_n$ to $\FFSM$ and gets the client share $z \in \{r,r+1\}$ of the $\FFSM$'s output (for a random masking value $r$). $\Sim$ proceeds in the following way to create the relevant columns of the garbled FSM in the internal simulation of the execution of protocol $\pi_{\mathsf{Helper}}$: 

\begin{enumerate}
    \item $\Sim$ chooses $n\cdot|Q|$ random keys of length $\kappa$ denoted by $k^i_q$, for $i\in\{1,\ldots,n\}, q\in\{1,\ldots,|Q|\}$.
    \item For $i=1,\ldots,n$, $\Sim$ samples an uniformly random value $q_i$ from $\{1,\ldots,|Q|\}$.
    \item For $i=1,\ldots,n-1$, $\Sim$ fixes 
\[
G_{\mathsf{fsm}}(q_i,x_i, i) :=
H(k_{q_i}^i || x_i) \oplus \left(q_{i+1}||  k^{i+1}_{q_{i+1}}\right),
\]
and it also fixes \[
G_{\mathsf{fsm}}(q_n,x_n, n) := H(k_{q_n}^n || x_n) \oplus z,
\]
so that the path that will be traversed by the client results in the desired share of the output.
    \item For $i=1,\ldots,n$, and all $q \in \{1,\ldots,|Q|\} \setminus \{q_i\}$, $\Sim$ fix 
\[
G_{\mathsf{fsm}}(q,x_i, i) := H(k_{q}^i || x_i) \oplus
\left( 0^{\kappa+\log|Q|} \right).
\]
    \item For $i=1,\ldots,n$, all $q \in \{1,\ldots,|Q|\}$ and all $\sigma \in \Sigma \setminus \{x_i\}$, $\Sim$ fix 
\[
G_{\mathsf{fsm}}(q,\sigma, i) := H(k_{q}^i || \sigma) \oplus
\left( 0^{\kappa+\log|Q|} \right).
\]

\end{enumerate}
Additionally, $\Sim$ generates the uniformly random key $k_{SH}$. Given the additive secret shares 
$x_{i,j}^S,x_{i,j}^H$ that the simulator $\Sim$ receives from $\Adv$ in the internal simulation of 
an execution of Protocol $\pi_{\mathsf{Helper}}$, $\Sim$ computes 
\[
G_{\mathsf{fsm}}^S(\cdot,x_i, i) \oplus H'(k_{SH} || i)
\]
and 
\[
G_{\mathsf{fsm}}^H(\cdot,x_i, i) \oplus H'(k_{SH} || i)
\] 
following the normal procedures used by the server and the helper, respectively, in Protocol $\pi_{\mathsf{Helper}}$ and sends those values to $\Adv$ as being the messages from the server and the helper. Additionally, $\Sim$ includes the key $k_{q_1}^1$ corresponding to the initial state in the simulated message from the server to $\Adv$. The simulator outputs the final view of the simulated adversary $\Adv$. 

The security against a semi-honest client follows trivially from the IND-CPA security of the symmetric encryption scheme that is used in the garbled transition matrices and in the XORing of the partial answers given by the server and the helper. A classical hybrid argument can be employed to replace one by one the encryptions of zero used by the simulator by the encryptions of the real garbled values used in the real world, thus proving that the real and ideal worlds are indistinguishable.

\textbf{Privacy against a Malicious Server:} The server only receives the secret shares from the client, but they look uniformly random from the server's point of view as long as the helper does not collude with the server. 

\textbf{Privacy against a Malicious Helper:} The helper receives the secret shares from the client, and the garbled transition matrices and $k_{SH}$ from the server. As long as the server does not collude with the helper, the secret shares received from the client look uniformly random from the helper's point of view. From the IND-CPA security of the symmetric encryption scheme that is used in the garbled transition matrices (and the fact that he does not know any state key as long as he does not collude with either the server or the client), the helper does not learn any information about the FSM.

\section{Conclusion}
In this paper, we presented a new protocol that strikes a better balance between the amount of communication and the number of rounds that are required. Furthermore, under the assumption of an untrusted third party we successfully achieve substantially low computation, communication, and round complexities.


\end{document}